\documentclass{article}
\usepackage{ijcai26}

\usepackage{times}
\usepackage{soul}
\usepackage{url}
\usepackage[hidelinks]{hyperref}
\usepackage[utf8]{inputenc}
\usepackage[small]{caption}
\usepackage{graphicx}
\usepackage{amsmath}
\usepackage{amsthm}
\usepackage{booktabs}
\usepackage{algorithm}
\usepackage{algorithmic}
\usepackage[switch]{lineno}

\usepackage{multirow}
\usepackage{makecell}
\usepackage{subcaption}
\usepackage{graphicx}

\usepackage{tcolorbox}
\usepackage{listings}
\usepackage{xcolor}
\usepackage[table]{xcolor} 
\usepackage{graphicx}
\usepackage{amsmath}

\definecolor{promptbg}{RGB}{248,248,248}
\definecolor{promptframe}{RGB}{210,210,210}
\lstdefinestyle{promptstyle}{
  basicstyle=\ttfamily\small,
  columns=fullflexible,
  breaklines=true,
  breakatwhitespace=true,
  keepspaces=true,
  showstringspaces=false,
  frame=none,
  tabsize=2,
  gobble=0,
  breakindent=0pt, 
  breakautoindent=false, 
  postbreak=\mbox{} 
}

\newtcolorbox{promptbox}[1][Prompt]{
  colback=promptbg,
  colframe=promptframe,
  title=\textbf{#1},
  boxrule=0.6pt,
  arc=1mm,
  coltitle=black,  
  colbacktitle=gray!25, 
  boxsep=4pt,  
  left=4pt, right=4pt, top=4pt, bottom=4pt
}

\title{Towards Automatic Evaluation and High-Quality
Pseudo-Parallel Dataset Construction for Audio
Editing: A Human-in-the-Loop Method}

\author{
Yuhang Jia$^1$$^*$,
Hui Wang$^1$ \thanks{Contribute equally. },
Xin Nie$^2$,
Yujie Guo$^1$,
Lianru Gao$^1$,
Yong Qin$^1$ \thanks{Corresponding author.}\\
\affiliations
$^1$College of Computer Science, Nankai University, Tianjin, China\\
$^2$College of Electronic Information and Optical Engineering, Nankai University, Tianjin, China\\
\emails
2013628@mail.nankai.edu.cn,
qinyong@nankai.edu.cn
}
\begin{document}

\maketitle

\begin{abstract}
Audio editing aims to manipulate audio content based on textual descriptions, supporting tasks such as adding, removing, or replacing audio events. Despite recent progress, the lack of high-quality benchmark datasets and comprehensive evaluation metrics remains a major challenge for both assessing audio editing quality and improving the task itself. In this work, we propose a novel approach for audio editing task by incorporating expert knowledge into both the evaluation and dataset construction processes: 1) First, we establish AuditScore, the first comprehensive dataset for subjective evaluation of audio editing, consisting of over 6,300 edited samples generated from 7 representative audio editing frameworks and 23 system configurations. Each sample is annotated by professional raters on three key aspects of audio editing quality: overall Quality, Relevance to editing intent, and Faithfulness to original features. 2) Based on this dataset, we systematically propose AuditEval, a family of automatic MOS-style evaluators tailored for audio editing, covering both SSL-based and LLM-based approaches. It addresses the lack of effective objective metrics and the prohibitive cost of subjective evaluation in this field. 3) We further leverage AuditEval to evaluate and filter a large amount of synthetically mixed editing pairs, mining a high-quality pseudo-parallel subset by selecting the most plausible samples. Comprehensive experiments validate that our expert-informed filtering strategy effectively yields higher-quality data, while also exposing the limitations of traditional objective metrics and the advantages of AuditEval. The dataset, codes and tools can be found at: {\url{https://github.com/NKU-HLT/AuditEval}}.
\end{abstract}

\section{Introduction}
Benefiting from the construction of Text-to-Audio (TTA) and Audio Caption (AC) datasets~\cite{kim2019audiocaps,gemmeke2017audio,drossos2020clotho,mei2024wavcaps}, audio generation has made significant strides in recent years. Current TTA models, including diffusion-based frameworks~\cite{huang2023make,liu2023audioldm,liu2024audioldm,ghosal2023text,chen2024musicldm,majumder2024tango,hung2024tangoflux,xue2024auffusion} and autoregressive transformer-based systems~\cite{agostinelli2023musiclm,kreukaudiogen,copet2023simple}, are capable of producing high-quality, diverse, and text-aligned audio outputs based on textual user prompts. However, beyond generation, audio editing remains equally essential but has received comparatively less attention. 
Unlike generation, which creates audio from scratch, audio editing modifies existing audio content based on textual instructions or target descriptions. It includes basic event manipulations (e.g., adding, removing, replacing) and more advanced tasks like super-resolution and style transfer. Audio editing faces three main challenges: maintaining the overall audio quality, executing edits that accurately reflect user intent, and preserving the unedited portions rather than generating a completely new sample.

Currently, the development of audio editing has been significantly hindered by the lack of high-quality paired datasets. This has also been observed in the image editing domain. Constructing supervised audio editing datasets is particularly difficult. Unlike TTA data, naturally occurring audio pairs that reflect specific editing operations are virtually nonexistent. Manual collection is often prohibitively expensive or infeasible, especially for scenarios like replacing gunshots in a crowd with thunder while preserving alignment.
As a result, most existing methods adopt training-free approaches based on generative diffusion models~\cite{ho2020denoising,song2020denoising} and inversion techniques~\cite{song2020denoising,mokady2023null,wu2023latent,huberman2024edit}. These methods repurpose the generation and cross-modal alignment capabilities of pretrained TTA models, steering generation toward desired edits by adjusting prompts or conditioning signals~\cite{paissan2023audio,xu2024prompt,zhang2024musicmagus,manor2024zero,jia2025audioeditor,liu2024medic,tsai2024audio}. However, they heavily depend on the underlying TTA model and often lack robustness due to the absence of task-specific supervision.

As an alternative to training-free approaches, supervised frameworks have also been explored by training models on explicitly constructed editing datasets. Audit~\cite{wang2023audit} represents the first supervised audio editing framework, introducing a triplet-based dataset and training a dedicated latent diffusion model (LDM). To date, it remains the only task-specific paired dataset for audio editing. This dataset is created through audio mixing and combinatorial pairing strategies. It leverages the compositional nature of audio, which unlike images, can be naturally mixed to simulate coherent edits.
A similar construction paradigm is adopted in InstructME~\cite{han2023instructme}, which applies mixing-based strategies to generate supervision signals for music editing tasks. Meanwhile, MusicGen~\cite{copet2023simple} and Instruct-MusicGen~\cite{zhang2024instruct} explore autoregressive approaches based on discrete audio-text tokens, supporting tasks such as continuation, transformation, and editing. WavCraft~\cite{liang2024wavcraft} further investigates instruction-based editing by employing large language models (LLMs) to interpret textual commands and call external audio and speech models for modification.

Despite these progress, there is still no unified standard for evaluating audio editing tasks. The absence of a widely accepted, open-source benchmark specifically designed for editing has led to fragmented evaluation protocols and a lack of reliable objective metrics. Currently, the most commonly used metric is the CLAP\_Score, which measures similarity between the edited audio and the target prompt in the embedding space, leveraging the CLAP model~\cite{elizalde2023clap} for cross-modal comparison without requiring ground-truth audio.
To assess preservation of unedited regions, some works~\cite{paissan2023audio,zhang2024musicmagus,jia2025audioeditor,liu2024medic} use the original audio as a reference and compute similarity between the original and edited versions. Others~\cite{xu2024prompt,zhang2024musicmagus} construct pseudo-paired datasets via audio mixing, following a strategy similar to that of Audit, and evaluate the generated outputs by comparing them with the mixed signals using conventional TTA metrics such as Fréchet Distance (FD), Fréchet Audio Distance (FAD)~\cite{roblek2019fr} and KL divergence.
However, such evaluation paradigms, which rely on synthetic mixtures as ground truth, often fall short. Because high-quality audio edits can vary substantially and are not necessarily aligned with a single reference.

As a result, subjective human evaluation remains the most reliable method for assessing audio editing quality. Current subjective evaluation protocols typically consider three aspects: overall audio \textbf{Quality}~\cite{zhang2024musicmagus,jia2025audioeditor}, \textbf{Relevance} to the editing intent~\cite{paissan2023audio,xu2024prompt,zhang2024musicmagus,manor2024zero,jia2025audioeditor,liu2024medic}, and \textbf{Faithfulness} to the original (unedited) content~\cite{paissan2023audio,xu2024prompt,zhang2024musicmagus,manor2024zero,jia2025audioeditor,liu2024medic}. These criteria provide an intuitive and comprehensive assessment of audio editing performance. Nevertheless, compared to pure generation tasks such as TTA, Text-to-Speech (TTS), or Text-to-Music (TTM), audio editing is inherently more complex, requiring deeper understanding and recognition of nuanced paired contents. This results in increased costs for annotator training, longer evaluation durations, and greater human labor requirements in practice. To these challenges, developing automatic Mean Opinion Score (MOS) prediction methods and frameworks for audio editing is extremely important and meaningful for advancing research in this field.

In this paper, we aim to bridge the data and evaluation gaps in audio editing by introducing AuditEval, a scalable automatic MOS-style assessment framework, and subsequently leveraging it to help mining  high-quality pseudo-parallel audio editing data. Our contributions are threefold:

\begin{itemize}
\item \textbf{Evaluation Dataset Construction}. We construct AuditScore, the first comprehensive dataset for subjective audio editing evaluation, consisting of over 6,300 edited pairs from 7+ representative frameworks and 23+ system variants. Each sample is annotated by professional raters from three key aspects: Quality, Relevance, and Faithfulness. 

\item \textbf{Automatic MOS-style Evaluation}. 
Based on AuditScore, we design and systematically study two paradigms for automatic MOS-style assessment in audio editing: 
a Self-Supervised Learning (SSL)-based evaluator (\textbf{\textit{AuditEval-ssl}}) and a Large Language Model (LLM)-based evaluator (\textbf{\textit{AuditEval-llm}}). We aim to provide scalable and cost-effective alternatives to expensive human evaluation.

\item \textbf{High-Quality Data Mining}. We further utilize AuditEval to assess and filter a large amount of synthetically mixed editing pairs, selecting only the most plausible samples to form a high-quality pseudo-parallel subset. Comprehensive experiments validate the effectiveness of this expert-informed filtering strategy in producing superior-quality data, while also highlighting the shortcomings of objective metrics and the advantages of AuditEval in audio editing task. 
\end{itemize}

\section{Related Work}
\subsection{Training-free Audio Editing Method}
Training-free audio editing methods exploit pre-trained TTA diffusion models to perform editing without the need for task-specific fine-tuning. These approaches typically begin by inverting the input audio into the noise space of the diffusion model, that is, mapping from $x_0$ (the observed audio) to $x_T$ (a noisy latent representation). The resulting latent $x_T$ preserves the semantic and structural features of the original audio while providing a flexible representation for guided generation. Editing is then achieved by denoising $x_T$ into a modified output $\hat{x}_0$, conditioned on user instructions or target descriptions.
Depending on the specific inversion strategy used, training-free audio editing methods can be broadly categorized into three types: SDEdit-based, DDIM-based, and DDPM-based approaches.

SDEdit-based methods perform audio editing by first applying a forward stochastic differential equation (SDE) process to produce a noised version of the original audio, and then conducting denoising under a modified prompt using the reverse SDE process (Eq.~\ref{eq:reverse_sde}). In this formulation, $\alpha(t)$ and $\sigma(t)$ are time-dependent functions that govern the noise injection level, and $\boldsymbol{s}_\theta$ represents the pretrained score model. The inversion timestep $t \in [0, T]$ plays a crucial role: larger $t$ values (closer to $T$) lead to more realistic yet less faithful edits, while smaller $t$ values better preserve the original content but reduce the editing strength.

\begin{equation}
\label{eq:reverse_sde}
\begin{aligned}
\mathbf{x_t} ={}& \mathbf{x_{t+\Delta t}} + \left(\textstyle\sigma^{2}(t)\!-\!\sigma^{2}(t+\Delta t)\right) \cdot \\
& \boldsymbol{s}_{\theta}(\mathbf{x_t}, t) + \sqrt{\textstyle\sigma^{2}(t)\!-\!\sigma^{2}(t+\Delta t)}\,\mathbf{z}
\end{aligned}
\end{equation}

DDIM-based approaches perform deterministic inversion of the diffusion process by exploiting the approximate reversibility of the underlying ODE with sufficiently small step sizes~\cite{ho2020denoising,song2020denoising,mokady2023null}. This allows for editing by first reversing the generation trajectory from $\mathbf{z}_0$ to $\mathbf{z}T$ (Eq.~\ref{eq:reverse_ddim}, where $\alpha$ is a hyperparameter and $\varepsilon{\theta}$ denotes the pretrained latent diffusion model), and then regenerating the audio using a modified prompt (Eq.~\ref{eq:forward_ddim}), thereby achieving semantically guided edits. Building upon DDIM inversion, several studies ~\cite{xu2024prompt,zhang2024musicmagus,jia2025audioeditor} incorporate enhancements such as attention manipulation and null-text optimization, achieving effective results in controllable audio editing.
\begin{equation}
\label{eq:reverse_ddim}
x_{t+1} = \textstyle \sqrt{\frac{\alpha_{t+1}}{\alpha_t}} x_t + \left(\sqrt{\frac{1 - \alpha_{t+1}}{\alpha_{t+1}}}\!-\!\sqrt{\frac{1 - \alpha_t}{\alpha_t}}\right) \cdot \varepsilon_{\theta}(x_t, t, \mathcal{P}_{\boldsymbol{ori}})
\end{equation}
\begin{equation}
\label{eq:forward_ddim}
x_{t-1} = \textstyle \sqrt{\frac{\alpha_{t-1}}{\alpha_t}} x_t + \left(\sqrt{\frac{1 - \alpha_{t-1}}{\alpha_{t-1}}}\!-\!\sqrt{\frac{1 - \alpha_t}{\alpha_t}}\right) \cdot \varepsilon_{\theta}(x_t, t, \mathcal{P}_{\boldsymbol{tar}})
\end{equation}

DDPM-based approaches include CycleDiffusion~\cite{wu2023latent} and Edit-Friendly DDPM~\cite{huberman2024edit}. This kind of method leverages the stochastic nature of the DDPM sampling process, specifically, the stochastic noise injected at each diffusion step to preserve the characteristics of the original input. By adding noise to the original input and storing intermediate latent states at each step, the model can compute the corresponding stochastic noise components (Eq.\ref{eq:ddpm}, $\mu_{t}\left(\mathrm{x}_{t}\right)$ is the calculated expectation of ${x}_{t}$, $\sigma_{t}$ is the hyperparameter), which are then reused during the denoising process under the modified prompt to retain original content. Edit-Friendly DDPM further improves by altering the noise injection mechanism so that the noise added at each timestep is independent, thereby enhancing editing controllability and performance. ZETA Editing~\cite{manor2024zero} is the first known method to apply DDPM inversion to audio editing, demonstrating superior results compared to SDEdit- and DDIM-inversion-based methods.
\begin{equation}
\label{eq:ddpm}
\mathbf{z}_{t}=\left(\mathrm{x}_{t-1}-\mu_{t}\left(\mathrm{x}_{t}\right)\right) / \sigma_{t}, \quad t=T, \ldots, 1
\end{equation}

\subsection{Supervised Audio Editing Models}

In contrast to training-free methods, supervised audio editing approaches rely on explicitly constructed editing datasets and can be broadly divided into non-autoregressive and autoregressive paradigms. Models such as Audit and InstructME employ latent diffusion models conditioned on audio-text multimodal inputs. They utilize specialized conditioning mechanisms and position-aware loss functions to enable end-to-end text-controlled audio editing~\cite{wang2023audit,han2023instructme}.
On the other hand, autoregressive approaches generate edited audio by conditioning on discrete representations of the original audio and textual edit instruction as context. Representative model like MusicGen adopt transformer-based decoders to sequentially synthesize edited audio ~\cite{copet2023simple,zhang2024instruct}. While these models benefit from strong language-audio alignment and high generation quality, they often struggle with temporal inconsistencies between the original and edited audio.

\vspace{-1mm}
\subsection{Auto-Evaluation for Speech, Music and Audio}

Automatic subjective evaluation plays a vital role in assessing AI-generated outputs by simulating human perception in a scalable and repeatable manner. Compared to objective metrics, it provides a more intuitive measure of perceptual quality. A well-designed evaluator not only aligns with human preferences but also guides the development of future models.
In the speech domain, automatic Mean Opinion Score (MOS) prediction has been widely adopted to evaluate tasks such as Text-to-Speech~\cite{cooper2022generalization,wang2023ramp}, Text-to-Voice~\cite{tang2024singmos}, and speech enhancement~\cite{reddy2021dnsmos,mittag2021nisqa,chen2022impairment}. These models have shown promising results and effectively complement  objective metrics.

In contrast, automatic evaluation for audio and music is more challenging. The content is highly diverse, ranging from natural sound events to complex musical structures. Moreover, the alignment between audio and textual descriptions is inherently looser than the tight correspondence found in speech transcription.
To address these challenges, MusicEval~\cite{liu2025musiceval} and AudioEval~\cite{wang2025audioeval} construct high-quality evaluation datasets and train scoring models to assess both overall quality and text relevance, achieving strong correlations with human judgments.
Meanwhile, Meta’s Audiobox~\cite{tjandra2025meta} proposes a unified framework for speech, music, and audio evaluation, defining four core aesthetic dimensions that collectively provide a comprehensive and structured standard for automatic audio assessment from an aesthetic perspective.

\section{Dataset Construction}
\subsection{Basic Information of AuditScore}

\begin{table*}[t]
  \centering
  \small
  \renewcommand{\arraystretch}{0.87}
  \setlength{\tabcolsep}{8pt}
  \begin{tabular}{ccccccc}
    \toprule
    \multirow{2}{*}{\textbf{\makecell[c]{Editing \\ Model or Method}}} & \multirow{2}{*}{\textbf{\makecell[c]{Generative \\ TTA Backbone}}} & \multirow{2}{*}{\textbf{Timesteps}} & \multirow{2}{*}{\textbf{System IDs}} & \multicolumn{3}{c}{\textbf{Average Editing Scores}} \\ \cmidrule(lr){5-7}
                      & & & & \textbf{\textit{Quality}} & \textbf{\textit{Relevance}} & \textbf{\textit{Faithfulness}} \\ \midrule
    \multirow{6}{*}{SDEdit} & \multirow{3}{*}{\textit{Tango2}} & 75 & Sys1 & 3.621\raisebox{-0.0ex}{\scriptsize$\pm$0.315} & 2.896\raisebox{-0.0ex}{\scriptsize$\pm$0.731} & 3.699\raisebox{-0.0ex}{\scriptsize$\pm$0.327} \\
                            & & 125 & Sys2 & 3.622\raisebox{-0.0ex}{\scriptsize$\pm$0.269} & 3.420\raisebox{-0.0ex}{\scriptsize$\pm$0.369} & 3.375\raisebox{-0.0ex}{\scriptsize$\pm$0.259} \\
                            & & 175 & Sys3 & 3.548\raisebox{-0.0ex}{\scriptsize$\pm$0.278} & 3.503\raisebox{-0.0ex}{\scriptsize$\pm$0.354} & 2.893\raisebox{-0.0ex}{\scriptsize$\pm$0.476} \\ \cmidrule(lr){2-7}
                            & \multirow{3}{*}{\textit{Auffusion}} & 75 & Sys4 & 3.602\raisebox{-0.0ex}{\scriptsize$\pm$0.284} & 3.268\raisebox{-0.0ex}{\scriptsize$\pm$0.354} & 3.426\raisebox{-0.0ex}{\scriptsize$\pm$0.476} \\
                            & & 125 & Sys5 & 3.487\raisebox{-0.0ex}{\scriptsize$\pm$0.287} & 3.461\raisebox{-0.0ex}{\scriptsize$\pm$0.386} & 2.885\raisebox{-0.0ex}{\scriptsize$\pm$0.486} \\
                            & & 175 & Sys6 & 3.492\raisebox{-0.0ex}{\scriptsize$\pm$0.303} & 3.531\raisebox{-0.0ex}{\scriptsize$\pm$0.349} & 2.742\raisebox{-0.0ex}{\scriptsize$\pm$0.418} \\ \midrule
    \multirow{2}{*}{DDIM Inversion} & \textit{Tango2} & 200 & Sys7 & 3.420\raisebox{-0.0ex}{\scriptsize$\pm$0.345} & 2.791\raisebox{-0.0ex}{\scriptsize$\pm$0.775} & 3.489\raisebox{-0.0ex}{\scriptsize$\pm$0.331} \\ \cmidrule(lr){2-7}
                                    & \textit{Auffusion} & 200 & Sys8 & 3.867\raisebox{-0.0ex}{\scriptsize$\pm$0.265} & 3.452\raisebox{-0.0ex}{\scriptsize$\pm$0.591} & 3.441\raisebox{-0.0ex}{\scriptsize$\pm$0.350} \\ \midrule
    \multirow{6}{*}{\makecell[c]{Edit-friendly \\ DDPM Inversion \\ (ZETA)}} & \multirow{3}{*}{\textit{Tango2}} & 75 & Sys9 & 3.658\raisebox{-0.0ex}{\scriptsize$\pm$0.238} & 3.330\raisebox{-0.0ex}{\scriptsize$\pm$0.338} & 3.641\raisebox{-0.0ex}{\scriptsize$\pm$0.257} \\
                            & & 125 & Sys10 & 3.684\raisebox{-0.0ex}{\scriptsize$\pm$0.274} & 3.354\raisebox{-0.0ex}{\scriptsize$\pm$0.455} & 3.479\raisebox{-0.0ex}{\scriptsize$\pm$0.301} \\
                            & & 175 & Sys11 & 3.655\raisebox{-0.0ex}{\scriptsize$\pm$0.242} & 3.593\raisebox{-0.0ex}{\scriptsize$\pm$0.281} & 3.408\raisebox{-0.0ex}{\scriptsize$\pm$0.256} \\ \cmidrule(lr){2-7}
                            & \multirow{3}{*}{\textit{Auffusion}} & 75 & Sys12 & 3.639\raisebox{-0.0ex}{\scriptsize$\pm$0.301} & 3.029\raisebox{-0.0ex}{\scriptsize$\pm$0.640} & 3.483\raisebox{-0.0ex}{\scriptsize$\pm$0.283} \\
                            & & 125 & Sys13 & 3.687\raisebox{-0.0ex}{\scriptsize$\pm$0.278} & 3.431\raisebox{-0.0ex}{\scriptsize$\pm$0.478} & 3.417\raisebox{-0.0ex}{\scriptsize$\pm$0.290} \\
                            & & 175 & Sys14 & 3.669\raisebox{-0.0ex}{\scriptsize$\pm$0.278} & 3.432\raisebox{-0.0ex}{\scriptsize$\pm$0.437} & 3.167\raisebox{-0.0ex}{\scriptsize$\pm$0.310} \\ \midrule
    \multirow{6}{*}{\makecell[c]{CycleDiffusion \\ DDPM Inversion}} & \multirow{3}{*}{\textit{Tango2}} & 75 & Sys15 & 3.592\raisebox{-0.0ex}{\scriptsize$\pm$0.280} & 2.848\raisebox{-0.0ex}{\scriptsize$\pm$0.632} & 3.595\raisebox{-0.0ex}{\scriptsize$\pm$0.258} \\
                            & & 125 & Sys16 & 3.445\raisebox{-0.0ex}{\scriptsize$\pm$0.317} & 3.184\raisebox{-0.0ex}{\scriptsize$\pm$0.544} & 3.313\raisebox{-0.0ex}{\scriptsize$\pm$0.291} \\
                            & & 175 & Sys17 & 3.293\raisebox{-0.0ex}{\scriptsize$\pm$0.265} & 3.146\raisebox{-0.0ex}{\scriptsize$\pm$0.393} & 3.054\raisebox{-0.0ex}{\scriptsize$\pm$0.344} \\ \cmidrule(lr){2-7}
                            & \multirow{3}{*}{\textit{Auffusion}} & 75 & Sys18 & 3.463\raisebox{-0.0ex}{\scriptsize$\pm$0.247} & 2.922\raisebox{-0.0ex}{\scriptsize$\pm$0.479} & 3.423\raisebox{-0.0ex}{\scriptsize$\pm$0.322} \\
                            & & 125 & Sys19 & 3.360\raisebox{-0.0ex}{\scriptsize$\pm$0.244} & 3.103\raisebox{-0.0ex}{\scriptsize$\pm$0.369} & 3.047\raisebox{-0.0ex}{\scriptsize$\pm$0.259} \\
                            & & 175 & Sys20 & 3.262\raisebox{-0.0ex}{\scriptsize$\pm$0.336} & 3.159\raisebox{-0.0ex}{\scriptsize$\pm$0.368} & 2.893\raisebox{-0.0ex}{\scriptsize$\pm$0.282} \\ \midrule
    AP-adapter & \textit{Audioldm2} & 200 & sys21 & 3.353\raisebox{-0.0ex}{\scriptsize$\pm$0.236} & 2.857\raisebox{-0.0ex}{\scriptsize$\pm$0.483} & 2.653\raisebox{-0.0ex}{\scriptsize$\pm$0.436} \\ \midrule
    Musicgen & \textit{Musicgen} & \textit{AutoRegressive} & sys22 & 3.208\raisebox{-0.0ex}{\scriptsize$\pm$0.230} & 2.968\raisebox{-0.0ex}{\scriptsize$\pm$0.575} & 2.693\raisebox{-0.0ex}{\scriptsize$\pm$0.447} \\ \midrule
    \textbf{Human Edit} & - & - & \textbf{sys23} & \textbf{3.883}\raisebox{-0.0ex}{\scriptsize$\pm$0.238} & \textbf{3.878}\raisebox{-0.0ex}{\scriptsize$\pm$0.412} & \textbf{3.988}\raisebox{-0.0ex}{\scriptsize$\pm$0.207} \\ \bottomrule
  \end{tabular}
  \caption{Subjective evaluation results across various systems. Detailed distribution of different editing types is presented in Appendix~\ref{appendixI}.}
  \vspace{-2mm}
  \label{table-sysid-score}
\end{table*}

AuditScore is a human-annotated dataset comprising 6,360 pairs of original and edited audio samples, totaling 35.3 hours in duration. These samples are generated using 23 distinct system configurations across seven representative audio editing approaches: SDEdit, DDIM, Edit-Friendly DDPM, CycleDiffusion, AP-adapter, MusicGen, and Human Edit(manual editing).
For diffusion-based methods (SDEdit, DDIM, Edit-Friendly DDPM, and CycleDiffusion), we adopt two state-of-the-art generative backbones Tango2 ~\cite{majumder2024tango} and Auffusion ~\cite{xue2024auffusion} to synthesize the edited audio. To further promote diversity and enhance model generalization, each configuration is instantiated with three inversion timesteps (75, 125, and 175 out of 200), allowing finer control over the degree of editing. This setup is specifically applied to SDEdit, Edit-Friendly DDPM, and CycleDiffusion to explore a wider range of editing intensities and semantic variations.

To construct a representative and diverse dataset, we first select original audio samples and corresponding captions from AudioCaps. The captions are clustered into 100 semantic groups using GTE-large~\cite{li2023towards} text embeddings, from which 2 to 3 audio samples are manually selected per cluster. For each selected audio, we design three types of editing instructions, including Addition, Deletion, and Replacement, resulting in a shared pool of 240 instructions across all systems. In addition, each representative systems are assigned 30 unique instructions to further increase the diversity of editing scenarios. Overall, AuditScore covers more than 200 distinct audio event types, with editing operations introducing over 150 novel events. The diversity in algorithms, model architectures, editing parameters, and audio content enables AuditScore to capture a wide range of editing behaviors and quality variations, making it a robust source of expert-informed knowledge for training and evaluating automatic audio editing evaluation models.

To assess the edited audio samples generated by these 23 systems, we conducted a large-scale subjective listening study involving \textbf{five domain experts} with extensive experience in audio dataset curation and perceptual evaluation. Each expert independently rated the editing audio pairs along three key perceptual dimensions: \textbf{Quality}, \textbf{Relevance}, and \textbf{Faithfulness}. The final rating for each utterance was computed as the average of five expert scores across individual dimensions. A comprehensive summary of system IDs and their evaluation results is presented in Table~\ref{table-sysid-score}.

\begin{figure*}[t!]
    \centering
\includegraphics[width=0.92\textwidth]{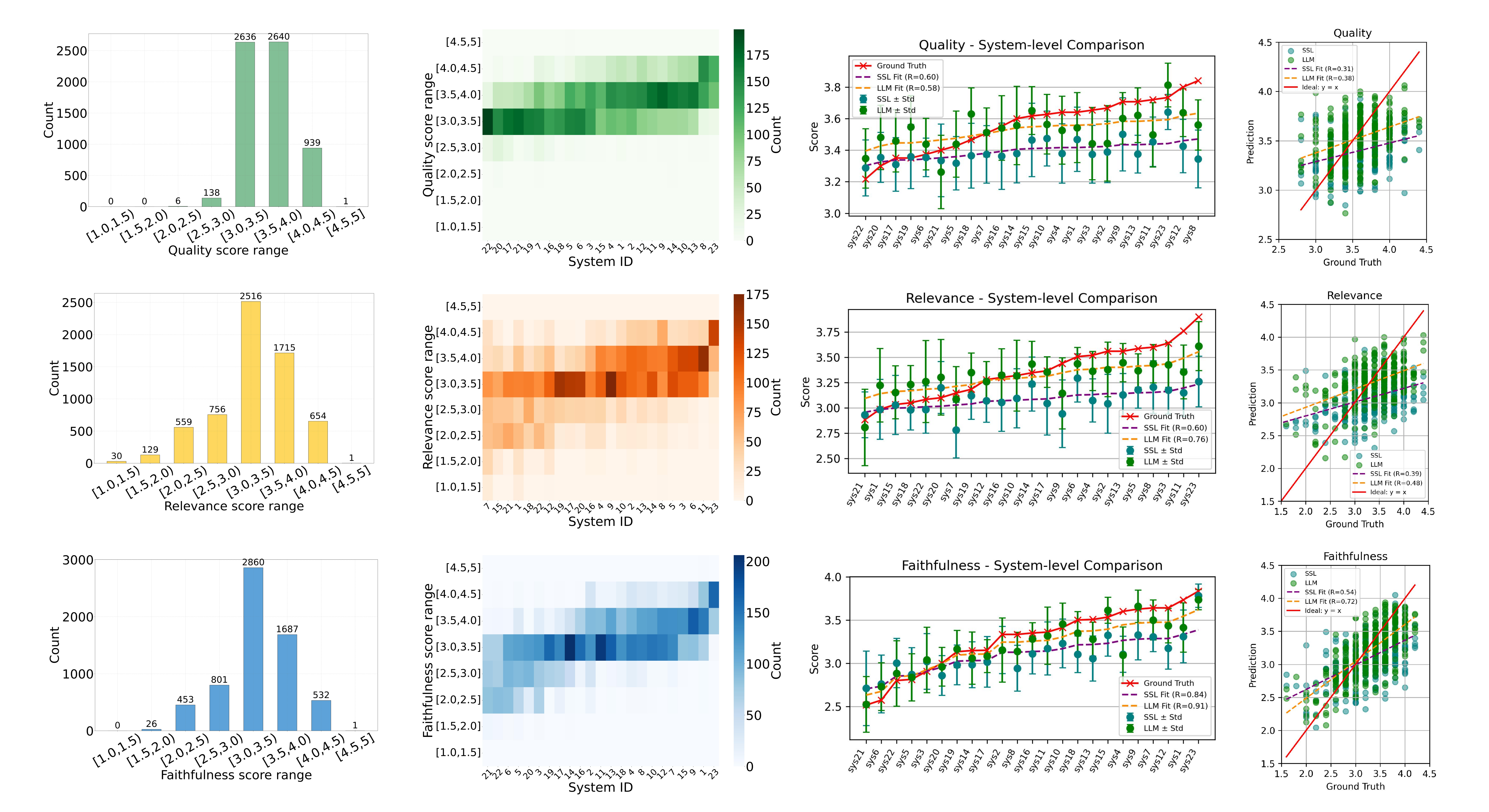}
    \caption{Distributions of AuditScore annotations (left) and AuditEval predictions (right) at both the system and utterance levels.}
    \vspace{-2mm}
    \label{fig:distributions}
\end{figure*}

\subsection{Scoring Protocol and Analysis}
To comprehensively evaluate audio editing performance, we adopt a three-dimensional scoring protocol encompassing Quality, Relevance, and Faithfulness. These dimensions are informed by established perceptual evaluation methodologies, such as MOS (Mean Opinion Score), CMOS (Comparison MOS)~\cite{loizou2011speech}, and semantic alignment assessments, and are further tailored to capture the specific nuances of audio editing. Detailed annotation guidelines were developed to ensure consistency across expert raters while preserving flexibility for nuanced perceptual judgments. The scoring scales are as follows(details can be found at Appendix~\ref{appendixA}):

\begin{itemize}
\item \textbf{Quality} —Assesses the overall semantic correctness and acoustic integrity of the edited audio:
(1) About the Same / Even Better, (2) Slightly Worse, (3) Noticeably Worse, (4) Much Worse, (5) Severely Degraded.

\item \textbf{Relevance} —Measures how well the edited audio semantically matches the editing instruction and target description:
(1) Excellent Match, (2) Good Match, (3) Partial Match, (4) Poor Match, (5) Mismatch.

\item \textbf{Faithfulness} —Evaluates the preservation of unedited content from the original audio in terms of rhythm, timing, loudness, and overall style:
(1) Perfectly Faithful, (2) Mostly Faithful, (3) Partially Faithful, (4) Minimally Faithful, (5) Not Faithful.
\end{itemize}

Figure~\ref{fig:distributions} the leftmost column, shows the distribution of scores across the three dimensions. The scores for Quality, Relevance, and Faithfulness each approximately follow a normal distribution, indicating that our scoring criteria provide a balanced and well-calibrated assessment scale. Notably, quality scores are slightly skewed toward higher values,\textit{ reflecting that most editing systems tend to maintain audio quality easily, while still facing challenges in precisely executing editing instructions and preserving unedited content.}

Figure~\ref{fig:distributions} the middle-left column, further visualizes performance across different systems. Notably, Human Editing consistently achieves the highest ratings in all dimensions, establishing an upper bound for automated models. Interestingly, some systems (e.g., sys3, sys5, and sys6) achieve high relevance scores, indicating strong alignment with the editing instructions, yet score lower in faithfulness, \textit{highlighting a common trade-off in current models between effective editing and content preservation}. This analysis underscores the remaining gap between human and machine editing and suggests a clear direction for future model improvement.

\section{Method}

\begin{figure*}[t!]
    \centering
\includegraphics[width=0.92\textwidth]{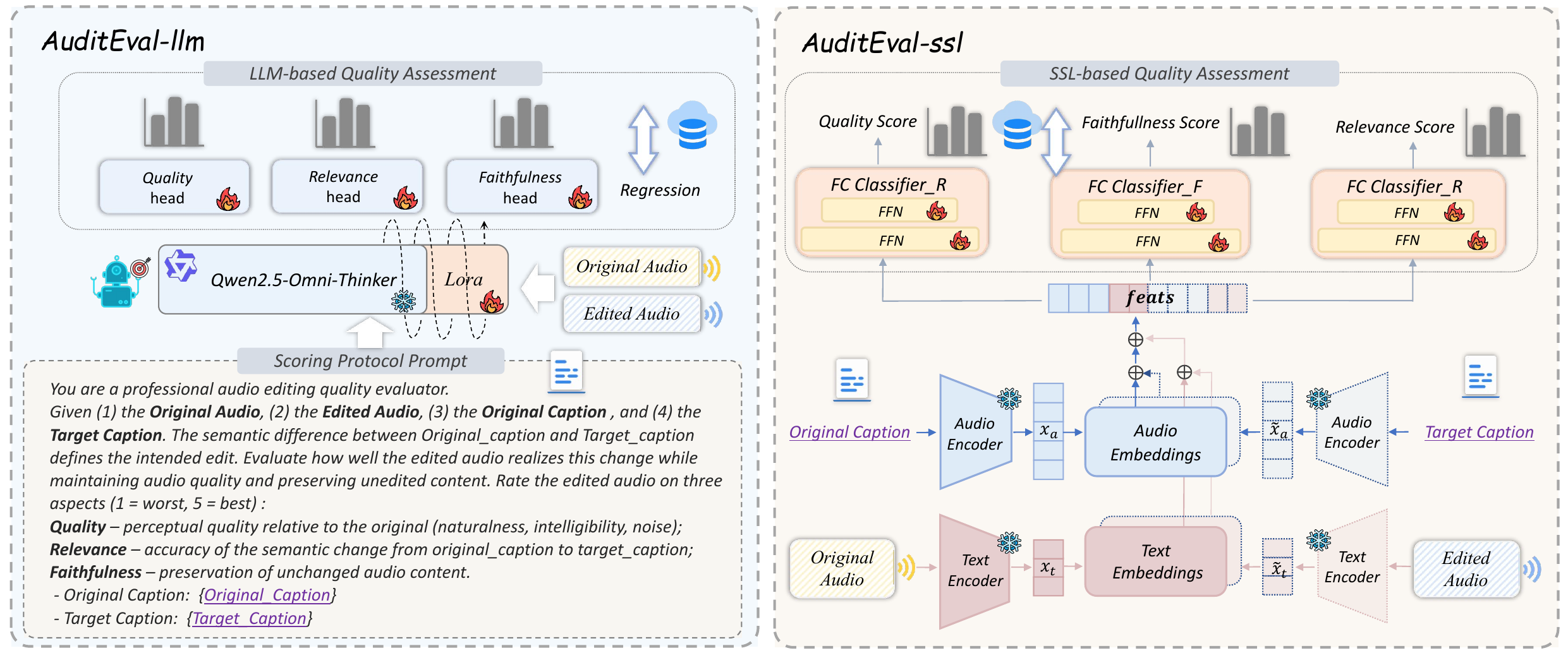}
    \caption{Model designs of \textit{\textbf{AuditEval-llm}} and \textit{\textbf{AuditEval-ssl}},
along with the corresponding training modules and inference workflows.}
    % \vspace{-3mm}}
    \label{fig:model}
\end{figure*}

\begin{table*}[t!]
  \centering
  \small
  \renewcommand{\arraystretch}{0.93}
  \setlength{\tabcolsep}{9.5pt}
  \begin{tabular}{c c cccc cccc}
    \toprule
    \multirow{2}{*}{\textbf{Dimensions}} 
    & \multirow{2}{*}{\textbf{Models}} 
    & \multicolumn{4}{c}{\textbf{System-level Prediction}} 
    & \multicolumn{4}{c}{\textbf{Utterance-level Prediction}} \\
    \cmidrule(lr){3-6} \cmidrule(lr){7-10}
    & 
    & MSE$\downarrow$ & LCC$\uparrow$ & SRCC$\uparrow$ & KATU$\uparrow$
    & MSE$\downarrow$ & LCC$\uparrow$ & SRCC$\uparrow$ & KATU$\uparrow$ \\
    \midrule
    \multirow{3}{*}{Quality}
    & \textit{Obj-Metric$^{*}$}
      & 8.7086 & 0.2277 & 0.2470 & 0.1858
      & - & - & - & -  \\
    & \textbf{AuditEval-ssl}
      & 0.0595 & \textbf{0.6101} & \textbf{0.7004} & \textbf{0.5516}
      & 0.1474  & 0.3318 & 0.3306 & 0.2462 \\
    & \textbf{AuditEval-llm}
      & \textbf{0.0198} & 0.5814 & 0.5523 & 0.3897
      & \textbf{0.0999} & \textbf{0.3840} & \textbf{0.3967} & \textbf{0.2922} \\
    \midrule
    \multirow{3}{*}{Relevance}
    & \textit{Obj-Metric$^{\dagger}$}
      & 8.7884 & 0.2141 & 0.1779 & 0.1225
      & 9.0698 & 0.1344 & 0.0812 & 0.0574 \\
    & \textbf{AuditEval-ssl}
      & 0.0962 & 0.6196 & 0.6472 & 0.4783
      & 0.2963 & 0.4105 & 0.3991 & 0.2886 \\
    & \textbf{AuditEval-llm}
      & \textbf{0.0328} & \textbf{0.7620} & \textbf{0.8142} & \textbf{0.6443}
      & \textbf{0.2189} & \textbf{0.4766} & \textbf{0.4049} & \textbf{0.2929} \\
    \midrule
    \multirow{3}{*}{Faithfulness}
    & \textit{Obj-Metric$^{\ddagger}$}
      & 6.7973 & 0.8790 & 0.8843 & 0.7024
      & 6.9466 & 0.4981 & 0.5038 & 0.3611 \\
    & \textbf{AuditEval-ssl}
      & 0.0726 & 0.8460 & 0.8809 & 0.7024
      & 0.2217 & 0.5327 & 0.5171 & 0.3791 \\
    & \textbf{AuditEval-llm}
      & \textbf{0.0285} & \textbf{0.9133} & \textbf{0.8952} & \textbf{0.7579}
      & \textbf{0.1313} & \textbf{0.7213} & \textbf{0.6898} & \textbf{0.5300} \\
    \bottomrule
  \end{tabular}
    \caption{Performance comparison of score prediction at both system and utterance levels.
        $^{*}$ denotes the Inception Score (IS), distribution-level metric for Quality,
        $^{\dagger}$ denotes the cross-modal CLAP score for Relevance,
        and $^{\ddagger}$ denotes the audio--audio CLAP score for Faithfulness.}
  \label{tab:result1}
\end{table*}

We present AuditEval, a scalable automatic MOS-style evaluation framework for audio editing that explores two complementary paradigms: an SSL-based predictor (\textit{AuditEval-ssl}) and an LLM-based evaluator (\textit{AuditEval-llm}), both producing aspect-wise scores for Quality, Relevance, and Faithfulness.

\subsection{AuditEval-SSL}
AuditEval-ssl is designed to automatically predict MOS-style scores for audio editing results by leveraging self-supervised audio representations. Given an original audio, an edited audio, and their corresponding original and target descriptions, the model first extracts high-level acoustic features and temporal cues from the audio using a pretrained SSL backbone (we adopt \textit{ms-clap}, which can automatically align audio and text semantics), and maps these representations to subjective scores along three evaluation dimensions. The core idea of AuditEval-ssl is to transfer the general-purpose audio understanding capability of SSL models to the audio editing evaluation setting, enabling robust and scalable scoring without relying on paired ground-truth edited audio.

\subsection{AuditEval-LLM} \label{sec::4.1}

AuditEval-llm explores an alternative evaluation paradigm in which a large language model serves as an evaluator by reasoning over multimodal evidence. Instead of directly regressing scores from latent embeddings, we convert the input audios together with their textual captions into a structured textual evidence set (details can be found at Appendix~\ref{appendixH}) and \textbf{prompt} the LLM to produce MOS-style ratings. In terms of model design, we adopt the advanced Large Audio Language Model \textit{Qwen2.5-Omni-7B}. Since the target scores are continuous float values obtained by averaging five independent human ratings, we move beyond conventional dialog-style discrete scoring in a \textit{\{1--5\}} set. Specifically, we retain the \textit{Qwen2.5-Omni Thinker} module and design three post regression heads to predict continuous scores for the three evaluation dimensions. This design leverages the instruction-following, cross-modal understanding, and judgment reasoning capabilities of LLMs for audio editing evaluation.

\subsection{Pseudo-parallel Data Filtering}
Due to the lack of well-established supervised datasets for audio editing, we construct a large-scale\textbf{Pseudo-parallel Audio Editing Dataset} following the methodologies of AUDIT~\cite{wang2023audit} and InstructME~\cite{han2023instructme}, and further enhance it using our automatic scoring model, AuditEval. The dataset consists of synthetic audio tuples (A, B, C, A+B, A+C), where Audio A is sampled from clustered AudioCaps~\cite{kim2019audiocaps} descriptions, and B/C are single-event clips paired with semantically similar prompts. These are reorganized into instruction-based triplets for \textbf{Addition}, \textbf{Deletion}, and \textbf{Replacement} tasks. A rule-based strategy is used to mix the audio and construct corresponding textual instructions, aiming to maintain acoustic coherence and semantic plausibility. Details are provided in Appendix~\ref{appendixB}.

However, this kind of rule-based concatenation introduces several limitations:
(1) the quality and prompt relevance of individual audio elements are inconsistent; (2) the insertion positions in the background audio are often suboptimal—even when attempting to insert at quieter regions, the resulting mixtures can sound unnatural and make it difficult to distinguish the editing change; (3) the randomly selected A/B/C combinations can be semantically implausible—for example, inserting the sound of paper rustling into a battlefield recording, or replacing male speech with a toilet flushing sound, etc.
To address these issues, we apply AuditEval to automatically evaluate and filter synthesized editing pairs based on scoring criteria informed by AuditScore. This filtering is expected to improve data quality and semantic plausibility, advancing toward a practical and scalable supervised audio editing dataset (detailed filtering configurations provided in Appendix~\ref{appendixC}).

\section{Experiments}
\label{section5}

\begin{table*}[t]
  \centering
  \small
  \renewcommand{\arraystretch}{0.94}
  \setlength{\tabcolsep}{5.0pt}
  \begin{tabular}{c ccccc ccccc}
    \toprule
    \multirow{2}{*}{\textbf{\makecell[c]{Sub-set Filtering \\ Strategy}}}
    & \multicolumn{5}{c}{\textbf{Rule-Based Objective Metrics}}
    & \multicolumn{5}{c}{\textbf{Human-Aligned Automatic Scores}} \\
    \cmidrule(lr){2-6} \cmidrule(lr){7-11}
     & Clap\_Score(\%) $\uparrow$
     & IS$\uparrow$
     & FAD$\downarrow$
     & FD$\downarrow$
     & KL$\downarrow$
     &  Human\_Clap(\%) $\uparrow$
     & PQ$\uparrow$
     & PC$\uparrow$
     & CE$\uparrow$
     & CU$\uparrow$ \\
    \midrule
    \textit{Random1}
     & 48.70 & \underline{12.23} & 0.2859 & 3.207 & 0.6913
     & 62.72 & 5.770 & 3.601 & 3.664 & 4.965 \\
    \textit{Random2}
     & 48.47 & \underline{12.23} & \textbf{0.2816} & \textbf{3.197} & 0.6892
     & 62.67 & 5.769 & 3.598 & 3.664 & 4.964 \\
    \textit{Random3}
     & 48.76 & 12.20 & \underline{0.2830} & \underline{3.205} & 0.6867
     & 62.74 & 5.769 & 3.601 & 3.667 & 4.965 \\
    \midrule
    \textbf{Filtered (AuditEval-ssl)}
     & \underline{49.43} & \textbf{12.61} & 0.4587 & 4.356 & \underline{0.6783}
     & \underline{64.51} & \textbf{5.835} & \underline{3.619} & \underline{3.775} & \textbf{5.105} \\
    \textbf{Filtered (AuditEval-llm)}
     & \textbf{50.96} & 11.86 & 0.6003 & 8.596 & \textbf{0.5289}
     & \textbf{64.98} & \underline{5.802} & \textbf{3.771} & \textbf{3.815} & \underline{5.051} \\
    \bottomrule
  \end{tabular}
 \caption{Comparison of pseudo-parallel data quality under different filtering strategies across objective metrics and human-aligned scores.}
  \vspace{-2mm}
  \label{tab:result2}
\end{table*}

\subsection{Experiment Setup}
\label{sec::5.1}
For MOS prediction, we train the proposed AuditEval family of models on the curated AuditScore dataset using a fixed 9:1 train–test split, with detailed training configurations provided in Appendix~\ref{appendixE}.
To evaluate consistency with human judgments, we report Mean Squared Error (MSE), Linear Correlation Coefficient (LCC), Spearman’s Rank Correlation Coefficient (SRCC), and Kendall’s Tau Correlation Coefficient (KTAU) at both the utterance and system levels, measuring prediction accuracy as well as rank-based agreement between predicted and ground-truth scores.

For the task of selecting high-quality pseudo-parallel audio editing data, we further conduct exploratory experiments combining objective and automatic subjective evaluations to compare editing quality before and after model-based filtering. In this setting, we retain the top 15\% of 59400 samples (over 3,300 hours) according to the predicted scores for comparison. This ratio is chosen as the minimal proportion at which the three random subsets exhibit comparable quality.

Regarding evaluation metrics, following practice in TTA, we adopt IS, FAD, FD, and KL divergence to assess similarity between audios~\cite{liu2023audioldm} before and after editing. For objective cross-modal evaluation, we use CLAP Score to measure alignment between edited audio and the corresponding target descriptions. For automatic subjective evaluation, we employ Human-CLAP~\cite{takano2025human}, which aligns CLAP-based scores with human preferences, as well as AudioBox to provide human-aligned aesthetic quality assessments of edited audio from four perceptual dimensions: Production Quality (\textbf{PQ}), Content Enjoyment (\textbf{CE}), Content Usefulness (\textbf{CU}), and Production Complexity (\textbf{PC}).

\subsection{Results and Analysis}

\paragraph{Effectiveness of AuditEval.}
As shown in Table~\ref{tab:result1} and the right columns of Fig.~\ref{fig:distributions}, both AuditEval-ssl and AuditEval-llm achieve practical and competitive performance in predicting audio editing quality. At the system level, which is commonly used to assess overall model performance and rank different systems, AuditEval-ssl attains a high LCC of around 0.6 on the Quality dimension, together with a low MSE below 0.1 on the 1--5 rating scale. Meanwhile, AuditEval-llm achieves remarkably high correlations of 0.76 and 0.91 on the Relevance and Faithfulness dimensions, respectively, while maintaining MSE values below 0.1, indicating substantially stronger alignment with human perception and clearly outperforming conventional objective metrics.

Compared with system-level evaluation, utterance-level prediction is inherently more challenging due to increased variability and noise at the individual sample level. Nevertheless, across all dimensions, both AuditEval-ssl and AuditEval-llm maintain relatively low MSE values (approximately 0.2 on the 1--5 scale) and preserve reasonable positive correlations (ranging from 0.38 to 0.72) with human judgments. In particular, AuditEval-llm achieves correlations of 0.47 for Relevance and 0.72 for Faithfulness, excelling in semantic related dimensions. These results lay a solid foundation at a finer granularity for subsequent downstream utterance-level data filtering and selection.

\paragraph{\textbf{AuditEval-ssl} vs. \textbf{AuditEval-llm}.}
A comparison between AuditEval-ssl and AuditEval-llm (Table~\ref{tab:result1} and Fig.~\ref{fig:distributions} right) shows that AuditEval-llm achieves state-of-the-art performance across most dimensions and metrics. In particular, on the Relevance and Faithfulness dimensions, which are closely related to semantic understanding, AuditEval-llm demonstrates notably stronger judgment capability. This advantage is likely attributable to the reasoning abilities and intelligence of large audio language models. From an acoustics-oriented perspective, AuditEval-ssl exhibits comparable, and in some metrics even superior, prediction performance on Quality dimension. This aligns with the general observation that audio-specific self-supervised learning models are particularly effective at extracting fine-grained acoustic representations, especially high-frequency cues that influence audio quality.

\paragraph{Effectiveness of Filtering.} As shown in Table~\ref{tab:result2}, the filtered data, selected using either AuditEval-ssl or AuditEval-llm, consistently exhibits improved performance across all human-aligned scores, as well as on CLAP\_Score, which is currently one of the most reliable objective metrics for audio editing. Notably, the subjective evaluation models used in our experiments, namely Human-CLAP and AudioBox, are entirely independent of AuditEval in terms of both data and model design. This fully in-the-wild evaluation setting provides strong evidence that the proposed filtering strategy leads to consistent and reliable quality improvements.

\paragraph{\textbf{Objective Metrics} vs. \textbf{Human-aligned Scores}.}
As shown in Table~\ref{tab:result2}, improvements in human-aligned scores are not consistently observed in objective metrics, contrary to expectations. Instead, metrics such as FAD and FD, which are commonly used to measure similarity between original and edited audios (faithfulness), exhibit noticeable degradation after filtering. This discrepancy reflects a fundamental difference between subjective and objective evaluation criteria. Human judgments tend to emphasize overall perceptual quality, including coherence, enjoyment, usability, and the semantic correctness of editing operations. In contrast, objective metrics such as FAD and FD primarily focus on low-level acoustic signal variations that are often perceptually insignificant to human listeners. These observations highlight the limitations of objective metrics and underscore the advantages of human-aligned evaluation, particularly for the complex and semantically rich audio editing scenarios.

\section{Conclusion}
In this paper, we address the long-standing challenges of data scarcity and unreliable evaluation in audio editing by grounding both evaluation and data construction in human judgments. We systematic construct \textit{AuditScore}, the first human-annotated dataset for MOS-style audio editing assessment, then develop \textit{AuditEval-ssl} and \textit{AuditEval-llm} to study two different paradigms for automatic subjective evaluation and high-quality dataset mining, providing both practical tools and methodological insights. Through comprehensive experiments, we reveal key challenges in audio editing evaluation and data curation, and highlight the role of human-aligned evaluation in improving evaluation and data quality.

\bibliographystyle{named}
\bibliography{ijcai26}

% \end{document}
\newpage
\appendix
\section{Details of Scoring Criteria Specification}
\label{appendixA}
Here, we present details about \textbf{Scoring Criteria Specification}. As described in the main text, we evaluate audio editing quality across three dimensions: \textbf{Quality}, \textbf{Relevance}, and \textbf{Faithfulness}. we establish detailed scoring criteria to ensure consistent and meaningful annotations. These guidelines integrate established evaluation practices including \textit{MOS} for perceived quality, \textit{CMOS} for relative assessment, and \textit{text-audio alignment} for semantic consistency. The detailed specifications for these three evaluation dimensions are presented below. Each dimension is scored on a 5-point scale (1-5), where higher values indicate better performance.

\textbf{Quality score} reflects how well the edited audio aligns with the editing instruction and target description, in terms of semantic correctness and execution: 
\begin{itemize}
    \item 5- About the Same/Even Better: The edited audio is on par with or even better than the original in terms of intelligibility, fluency, and background noise.
    \item 4- Slightly Worse: At most one aspect (e.g., mild mechanical artifact or minor increase in noise) is slightly degraded, but listening quality remains unaffected.
    \item 3- Noticeably Worse: At least two aspects show moderate degradation, or one aspect is clearly degraded. Artifacts such as robotic tone, discontinuities, or increased noise may affect understanding in some segments but remain tolerable.
    \item 2- Much Worse: Significant degradation in at least two aspects, with noticeable artifacts severely affecting naturalness and intelligibility.
    \item 1- Badly Damaged: Severe issues across all aspects, such as overwhelming noise, harsh distortions, or unintelligible content, rendering the audio difficult to comprehend.
\end{itemize}

\textbf{Relevance score} reflects how well the edited audio aligns with the editing instruction and target description, in terms of semantic correctness and execution:
\begin{itemize}
    \item 5- Excellent Match: The edit perfectly matches the described intent. For Addition, the added event feels natural and contextually appropriate. For Deletion, the targeted event is successfully removed without leaving any noticeable residue. For Replacement, the substituted content is fluent and coherent, maintaining the overall auditory experience.
    \item 4- Good Match: Overall alignment is correct, but there may be minor imperfections, such as slightly abrupt additions or faint remnants after deletion.
    \item 3- Partial Match: The general category or type of edit is correct, but details may be awkward or somewhat misaligned with the description.
    \item 2- Poor Match: The edit loosely corresponds to the intent, but contains semantic inaccuracies or unnatural execution that hampers comprehension.
    \item 1- Mismatch: The edited content fails to reflect the described intent, with missing, incorrect, or misleading modifications.
\end{itemize}

\textbf{Faithfulness criterion} evaluates how well the unedited portions of the audio are preserved in terms of rhythm, sequence, loudness, and overall style:

\begin{itemize}
    \item 5- Perfectly Faithful: Unedited segments are completely preserved. Prosody, order, and audio style remain fully consistent with the original.
    \item 4- Mostly Faithful: Slight variation in loudness or subtle stylistic shifts may occur, but prosody and content are largely preserved.
    \item 3- Partially Faithful: Noticeable changes in rhythm or sequence are present, though the overall style remains similar. Minor alterations in the unedited content may be perceptible.
    \item 2- Low Faithfulness: Major changes in prosody, ordering, or style, with only a few elements traceable to the original.
    \item 1- Not Faithful: The edited audio bears little or no resemblance to the original; unedited content is altered significantly or missing altogether.
\end{itemize}

\section{Details of Pseudo-Parallel Audio Editing Dataset Construction.}
\label{appendixB}
Inspired by AUDIT and InstructME, we also adopt an audio mixing strategy to construct our \textbf{Pseudo-Parallel Audio Editing Dataset}. The goal of the dataset construction process is to generate a set of audio pairs comprising (A, B, C, A + B, A + C), which serves as the foundation for subsequent audio editing tasks. For audio types A, we collected over 45,000 audio prompts from the AudioCaps dataset~\cite{kim2019audiocaps}. For audio types B and C, we select prompts corresponding to common single-audio events, covering eight broad categories—NaturalSounds, Transportation, IndoorActivities, OutdoorActivities, Animals, Music, Speech and HumanSounds—spanning over 100 distinct sound types. By pairing these prompts in all possible two-way combinations, we obtained 4950 pairs of audio prompts. Then, for each audio prompt pair (B, C), we randomly select 20 audio prompts from the 45,000+ A audio prompts to form combinations, denoted as (A, B, C).

We utilize TangoFlux~\cite{hung2024tangoflux} to generate audio based on the (A, B, C) prompts, using a step of 50 and a guide scale of 4.5. The generation rules are as follows:
\begin{itemize}
    
\item \textbf{Audio A:} The generated audio has a fixed length of 10 seconds.

\item \textbf{Audio B:} The duration of the generated audio is randomized, with a duration ranging from 3 to 6 seconds.

\item \textbf{Audio C:} The duration of the generated audio matches that of the audio B.
 
\end{itemize}
Then we present the method used to mix the audio A, B, and C to obtain the audio A+B and A+C. Taking the generation of A+B as an example, let the duration of audio B be denoted as $t_B$. We apply a sliding window approach to identify the segment in audio A with the lowest energy and a length equal to $t_B$, which is then selected as the insertion interval for audio B. To reduce the search time, the stride of the sliding window was set to 0.1 seconds. To ensure consistency of volume and alignment of the dynamic range between the two audio segments, the amplitude of audio B is adjusted. Specifically, peak amplitude normalization is applied to audio B so that its maximum amplitude matches that of audio A, preventing audio B from being either masked by or excessively prominent compared to audio A.

\begin{table}[h!]
\centering
\caption{Rule-based template for generating A+B audio prompts}
\label{tab:rule_based_template}
\renewcommand{\arraystretch}{1.2}
\setlength{\tabcolsep}{4pt}
\begin{tabular}{c|c|l}
\hline
Condition & Probability & Sentence Template \\
\hline
\multirow{7}{*}{$c_i < split_1$}
& 30\% & $P_B$, $P_A$. \\
& 20\% & $P_B$, followed by $P_A$. \\
& 20\% & $P_B$, then $P_A$. \\
& 10\% & with $P_B$, $P_A$. \\
& 10\% & $P_B$, and $P_A$. \\
& 5\%  & After $P_B$, $P_A$. \\
& 5\%  & $P_B$ before $P_A$. \\
\hline
\multirow{7}{*}{$c_i > split_2$}
& 30\% & $P_A$, $P_B$. \\
& 20\% & $P_A$, followed by $P_B$. \\
& 20\% & $P_A$, then $P_B$. \\
& 10\% & $P_A$, with $P_B$. \\
& 10\% & $P_A$, and $P_B$. \\
& 5\%  & After $P_A$, $P_B$. \\
& 5\%  & $P_A$ before $P_B$. \\
\hline
\multirow{4}{*}{$split_1 \leq c_i \leq split_2$}
& 30\% & $P_A$, with $P_B$. \\
& 30\% & $P_A$, while $P_B$. \\
& 20\% & $P_A$, $P_B$. \\
& 20\% & $P_A$, and $P_B$. \\
\hline
\end{tabular}
\end{table}

After completing the audio insertion, we proceed to generate the corresponding audio prompts for A+B and A+C. A rule-based template method is adopted for prompt generation.Taking the generation of the A+B audio prompt as an example, we define two segmentation points, $split_1$ and $split_2$(with $0 < split_1 < split_2 < 10$). Audio A is then divided into three segments:$[0,split_1)$, $[split_1,split_2]$, and $(split_2,10]$. Let $c_i$denote the center point of the inserted audio B. Let $prompt_A$ be the textual description of audio A, and $prompt_B$ that of audio B. The rules for generating the composite audio prompt are summarized in the table \ref{tab:rule_based_template}. The generation rules for the audio prompt corresponding to A+C are identical to those used for A+B. For a 10-second audio segment, we use a $split_1$ of 3 and a $split_2$ of 7. In practice, we observe that the insertion position tends to satisfy either $c_i < split_1$ or $c_i > split_2$, which may be attributed to the inherent generation characteristics of TangoFlux. To ensure a roughly uniform probability of $c_i$ falling into each of the three segments, we introduce a \textbf{\textit{drop}} operation: a portion of the center points $c_i$ that meet the above two conditions are reassigned to a random value within the interval $[split_1,split_2]$.

Following AUDIT, we reorganize the tuples (A, B, C, A + B, A + C) into the format of (instruction, input audio, output audio), which servers as training data for audio editing tasks. Specifically, (Add, A, A + B) and (Add, A, A + C) constitute two training examples for the \textbf{\textit{add}} task; (Delete, A + B, A) and (Delete, A + C, A) constitute two training examples for the \textbf{\textit{delete}} task; (Replace, A + B, A + C) and (Replace, A + C, A + B) constitute two training examples for the \textbf{\textit{replace}} task.

\begin{figure*}[t!]
    \centering
\includegraphics[width=0.80\textwidth]{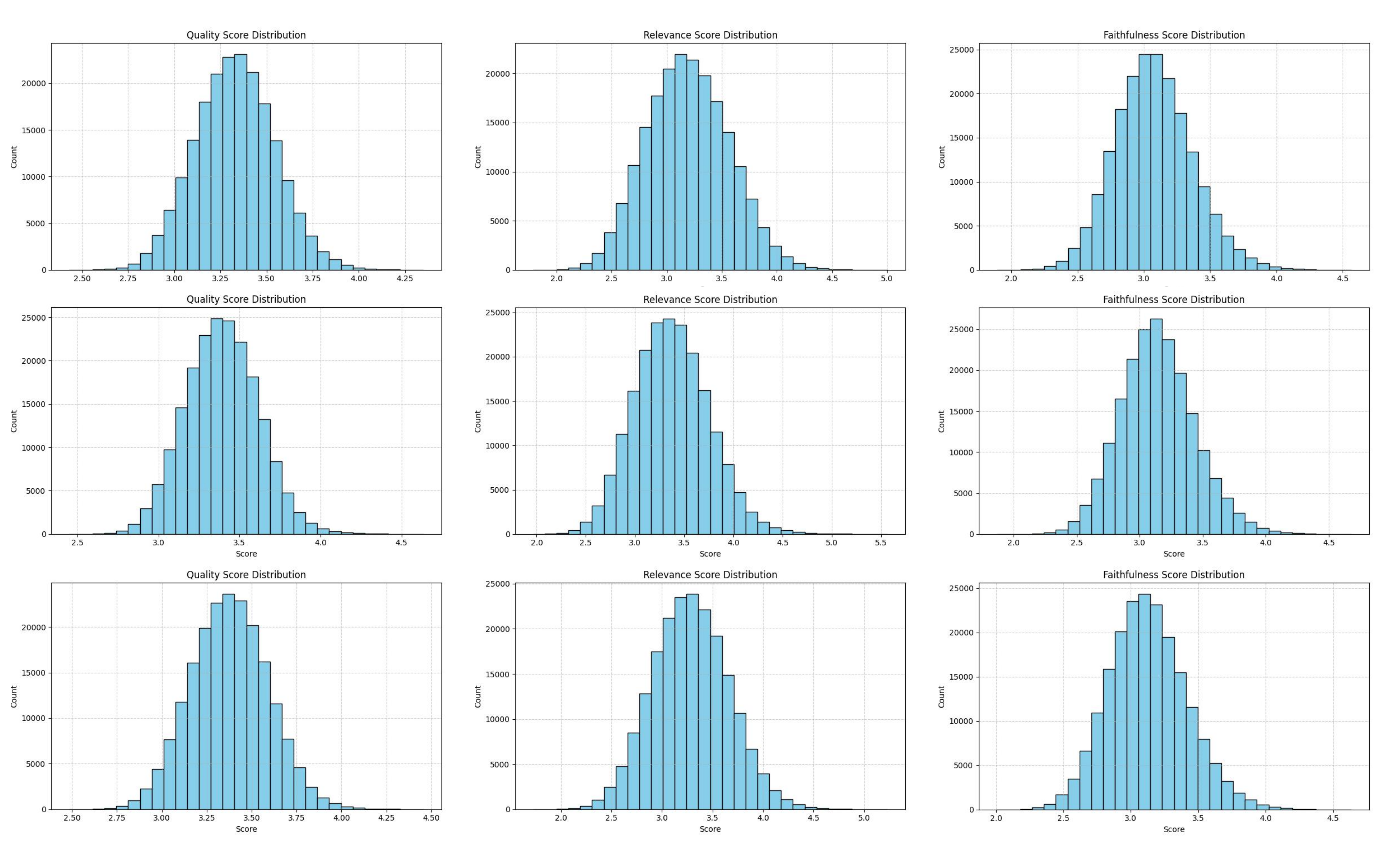}
    \caption{Score prediction distributions on the large-scale Pseudo-Parallel Audio Editing Dataset, with addition, deletion, and modification operations shown from top to bottom.}
    \label{fig:filter_ssl}
\end{figure*}
\begin{figure*}[t!]
    \centering
\includegraphics[width=0.80\textwidth]{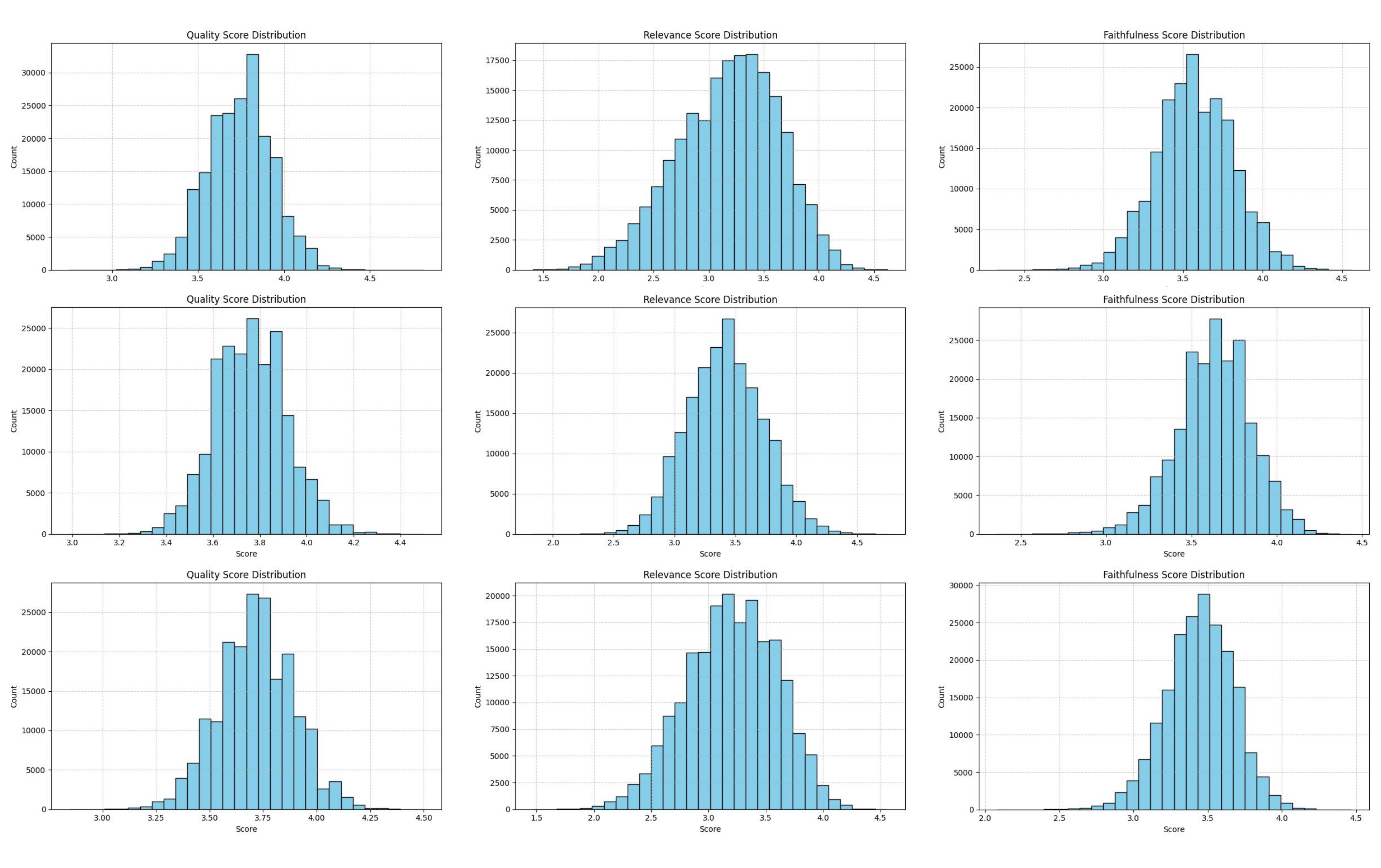}
    \caption{Score prediction distributions on the large-scale Pseudo-Parallel Audio Editing Dataset, with addition, deletion, and modification operations shown from top to bottom.}
    \label{fig:filter_llm}
\end{figure*}

\section{Filtering Strategy of High-Quality Pseudo-parallel Audio Editing Dataset}\label{appendixC}

\begin{table}[h!]
\centering
\renewcommand{\arraystretch}{1.2}
\caption{Number of Samples After Each Round of Selection(ssl)}
\label{tab:after_selection_ssl}
\begin{tabular}{lcccc}
\hline
Selection Stage & Add & Delete & Replace \\
\hline
Original number  & 198000 & 198000 & 198000 \\
After Round 1    & 59400  & 59400  & 59400  \\
After Round 2    & 40509  & 48091  & 45457  \\
After Round 3    & 30000  & 30000  & 30000  \\
\hline
\end{tabular}
\end{table}

\begin{table}[h!]
\centering
\renewcommand{\arraystretch}{1.2}
\caption{Number of Samples After Each Round of Selection(llm)}
\label{tab:after_selection_llm}
\begin{tabular}{lcccc}
\hline
Selection Stage & Add & Delete & Replace \\
\hline
Original number  & 198000 & 198000 & 198000 \\
After Round 1    & 99000  & 99000  & 99000  \\
After Round 2    & 68551  & 76358  & 68522  \\
After Round 3    & 30000  & 30000  & 30000  \\
\hline
\end{tabular}
\end{table}

Here, we present the filtering strategy of constructing \textbf{High-Quality Pseudo-Parallel Audio Editing Dataset}. In AppendixC, the construction of the Pseudo-Parallel Audio Editing Dataset is presented, comprising 30,000 training instances for each of the three editing tasks: Addition, Deletion, and Replacement. We employ our proposed model, AuditEval, to evaluate the dataset across three dimensions, with the score distributions illustrated in Fig.~\ref{fig:filter_ssl} and Fig.~\ref{fig:filter_llm}. Subsequently, based on the scores of these three dimensions, we perform three rounds of filtering on the dataset. The filtering criteria are as follows:
\begin{itemize}
\item \textbf{Stage 1: Faithfulness-based filtering.}
All samples are sorted by their predicted faithfulness scores in descending order. 

\item \textbf{Stage 2: Quality thresholding.}
Among the retained samples, we further filter out those with predicted quality scores boardline.

\item \textbf{Stage 3: Relevance-based selection.}
The remaining samples are ranked by predicted relevance scores, and the top 30,000 samples are selected to form the final dataset.
\end{itemize}

For AuditEval-ssl, we adopt a quality threshold of 3.40, faithfulness thresholds of 3.216 / 3.281 / 3.266, and relevance thresholds of 3.159 / 3.484 / 3.344 for the three successive filtering rounds.
For AuditEval-llm, the corresponding thresholds are set to 3.70 for quality, 3.578 / 3.641 / 3.453 for faithfulness, and 3.500 / 3.609 / 3.469 for relevance.

Tables~\ref{tab:after_selection_ssl} and~\ref{tab:after_selection_llm} illustrate the multi-round selection process for the two evaluators, respectively. In each round, the selected thresholds are designed to remove approximately half of the remaining samples, progressively retaining higher-quality pseudo-parallel data.
 
\section{User Interface for Subjective Evaluation}
\label{appendixD}

\begin{figure*}[t!]
    \centering
    \includegraphics[width=0.95\textwidth]{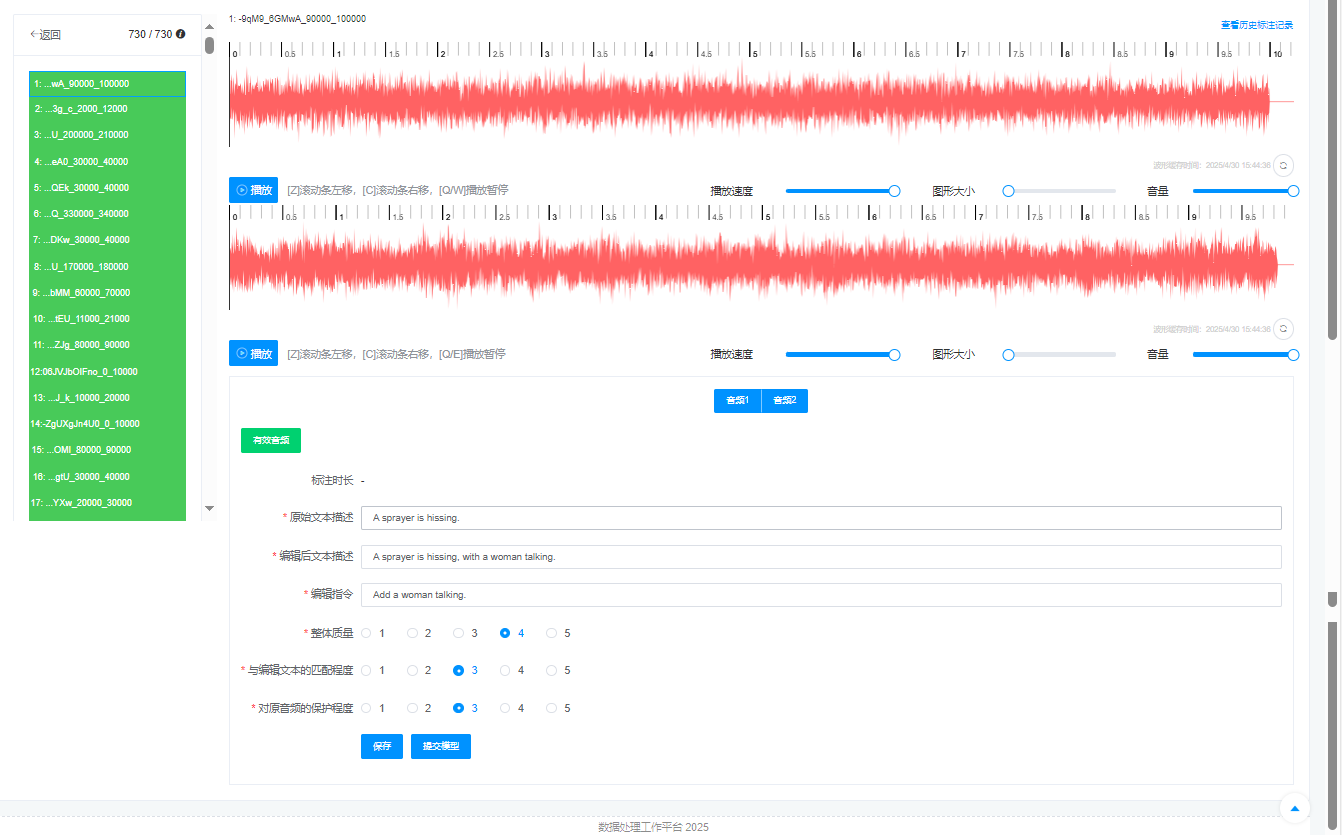}
    \caption{User Interface for Subjective Evaluation.}
    \label{fig:UI}
\end{figure*}

To ensure consistency and ease of use during the manual scoring process, we designed a custom web-based interface for subjective evaluation. As shown in Figure~\ref{fig:UI}, the interface presents the original audio, the edited audio, the original prompt, the edited prompt, and, if available, the edit instruction. Annotators are instructed to score each sample along three dimensions: Quality, Relevance, and Faithfulness, following the same evaluation criteria defined in the AuditScore dataset. This standardized interface streamlines the annotation workflow and helps reduce variability in human ratings.

\section{Computational Experiments Details}
\label{appendixE}

All experiments were conducted on a single NVIDIA H800 GPU. To ensure reproducibility, we fixed all sources of randomness with a global random seed set to \texttt{1984}.

The AuditEval-ssl model was trained using the Adam optimizer with a learning rate of $5 \times 10^{-5}$. We employed a \texttt{ReduceLROnPlateau} scheduler with \texttt{mode='min'}, \texttt{factor=0.8}, and \texttt{patience=5}. Training was conducted with a batch size of 256 for training and 64 for validation. The model comprises three independent regression heads for Quality, Relevance, and Faithfulness, which were trained for 50, 30, and 35 epochs, respectively.

AuditEval-llm was trained within the ms-swift framework as a regression-based sequence classification model that predicts three continuous scores corresponding to Quality, Relevance, and Faithfulness (\texttt{num\_labels=3}, \texttt{problem\_type=regression}), with chat templates disabled to ensure a unified input format. We adopted LoRA-based parameter-efficient fine-tuning, applying LoRA to all linear layers with a rank of 8 and a scaling factor (\texttt{alpha}) of 32, while freezing the ViT backbone. The model was trained for 6 epochs using the Adam optimizer with a learning rate of $1 \times 10^{-4}$ and a warmup ratio of 0.05, under mixed-precision training with \texttt{bfloat16}. The per-device batch size was set to 8 for training and 2 for evaluation, with gradient accumulation over 8 steps and a maximum input sequence length of 4096 tokens.

\section{Licenses for assets}\label{appendixF}
All assets developed in this work, including datasets, checkpoints, training protocols, and code, will be released under the CC BY 4.0 License.

\section{Ethics Statement}\label{appendixG}
We employed professional annotators via a third-party commercial provider to ensure consistent and reliable subjective evaluations. Annotators were fairly compensated above local minimum wage in accordance with regional labor laws. They received clear instructions on task objectives, scoring criteria, and example cases prior to annotation. No personal or sensitive information was collected, and no individual-level data beyond task performance was recorded. 

LLMs like GPT-5 were used appropriately in this work, primarily for language polishing. We ensured that all content was human-authored and carefully reviewed, with no reliance on LLMs for generating scientific claims or experimental results, thereby avoiding potential AI hallucinations.

\section{Details of the AuditEval-llm Prompt}\label{appendixH}
Figure~\ref{fig:example_prompt1} and Figure~\ref{fig:example_prompt2} illustrate the prompts used in AuditEval-llm, which instruct the model to act as a professional audio editing quality evaluator.
\label{appendixE}
\begin{figure}[h!]
    \centering
    \begin{promptbox}[Prompt]
    \begin{lstlisting}[style=promptstyle]
You are a professional audio editing quality evaluator.

You will be given:
(1) the original audio,
(2) the edited audio,
(3) a textual description of the original audio (original_caption),
(4) a textual description of the target edited audio (target_caption).

The semantic difference between original_caption and target_caption implicitly defines the intended audio edit.
Your task is to evaluate how well the edited audio reflects this semantic change,
while maintaining high audio quality and preserving unedited content.
Evaluate the edited audio from three perspectives: Quality, Relevance, and Faithfulness.
Each score should be an integer from 1 to 5 (higher is better).
Output only three integer scores, without explanations.

1. Quality (1-5):
Evaluate the perceptual quality of the edited audio compared to the original,
including intelligibility, fluency, naturalness, and background noise.

- 5: As good as or better than the original, with clear, fluent, and natural audio.
- 4: Slight degradation in at most one aspect, but overall quality remains acceptable.
- 3: Noticeable degradation in multiple aspects or clear degradation in one aspect, but still understandable.
    \end{lstlisting}
    \end{promptbox}
    \caption{Examples AuditEval-llm Promps(I)}
    \label{fig:example_prompt1}
\end{figure}

\begin{figure}[ht]
    \centering
    \begin{promptbox}[Prompt]
    \begin{lstlisting}[style=promptstyle]
- 2: Significant artifacts that seriously harm naturalness or intelligibility.
- 1: Severely damaged or unintelligible audio.

2. Relevance (1-5):
Evaluate how well the edited audio matches the semantic content described in target_caption,
relative to original_caption.
Focus only on the parts that are different between original_caption and target_caption.

- 5: The edited audio accurately and naturally reflects the target_caption, with correct and complete semantic changes.
- 4: The main semantic change is correctly reflected, with only minor imperfections.
- 3: The general semantic intent is correct, but details are partially misaligned or awkward.
- 2: The semantic change is loosely reflected, with clear errors or unnatural execution.
- 1: The edited audio does not match the target_caption.

3. Faithfulness (1-5):
Evaluate how well the audio content that should remain unchanged (according to the captions)
is preserved in the edited audio, in terms of rhythm, order, loudness, prosody, and style.
Do not penalize the semantic changes required by target_caption.

- 5: Unchanged content is perfectly preserved and indistinguishable from the original.
- 4: Mostly preserved, with only minor variation.
- 3: Noticeable changes in rhythm or prosody, but overall style remains similar.
- 2: Major changes in unedited content with limited resemblance.
- 1: Unedited content is heavily altered or missing.

Original Caption:
{original_caption}

Target Caption:
{target_caption}
    \end{lstlisting}
    \end{promptbox}
    \caption{Examples AuditEval-llm Promps(II)}
    \label{fig:example_prompt2}
\end{figure}

\section{Evaluation scores Across Editing Operations (Add / Delete / Replace)}\label{appendixI}
This section presents a breakdown of evaluation score across different audio editing operations, including \textbf{Add}, \textbf{Delete}, and \textbf{Replace}, as shown in following table:
\begin{table*}[h!]
\centering
% \caption{Scores by Editing Operation (Add / Delete / Replace).}
\renewcommand{\arraystretch}{0.92}
\setlength{\tabcolsep}{4pt}
\begin{tabular}{cccccccc}
\hline
Algorithm & TTA backbone & Timesteps & System ID & Operation & Quality score & Relevance score & Faithfulness score \\ \hline
\multirow{18}{*}{SDEdit} & \multirow{9}{*}{Tango2} & \multirow{3}{*}{75} & \multirow{3}{*}{sys1} & Add & 3.616 & 3.206 & 3.646 \\
& & & & Del & 3.560 & 2.466 & 3.726 \\
& & & & Rep & 3.686 & 3.016 & 3.726 \\ \cline{3-8}
& & \multirow{3}{*}{125} & \multirow{3}{*}{sys2} & Add & 3.656 & 3.550 & 3.480 \\
& & & & Del & 3.602 & 3.314 & 3.420 \\
& & & & Rep & 3.608 & 3.396 & 3.226 \\ \cline{3-8}
& & \multirow{3}{*}{175} & \multirow{3}{*}{sys3} & Add & 3.562 & 3.498 & 3.108 \\
& & & & Del & 3.478 & 3.504 & 2.896 \\
& & & & Rep & 3.604 & 3.508 & 2.674 \\ \cline{2-8}
& \multirow{9}{*}{Auffusion} & \multirow{3}{*}{75} & \multirow{3}{*}{sys4} & Add & 3.704 & 3.356 & 3.602 \\
& & & & Del & 3.552 & 3.122 & 3.418 \\
& & & & Rep & 3.550 & 3.326 & 3.258 \\ \cline{3-8}
& & \multirow{3}{*}{125} & \multirow{3}{*}{sys5} & Add & 3.518 & 3.408 & 3.228 \\
& & & & Del & 3.412 & 3.458 & 2.728 \\
& & & & Rep & 3.530 & 3.518 & 2.700 \\ \cline{3-8}
& & \multirow{3}{*}{175} & \multirow{3}{*}{sys6} & Add & 3.502 & 3.442 & 2.826 \\
& & & & Del & 3.494 & 3.650 & 2.724 \\
& & & & Rep & 3.480 & 3.500 & 2.676 \\ \hline
\multirow{6}{*}{DDIM-Inv} & \multirow{3}{*}{Tango2} & \multirow{3}{*}{200} & \multirow{3}{*}{sys7} & Add & 3.522 & 2.738 & 3.604 \\
& & & & Del & 3.358 & 2.840 & 3.462 \\
& & & & Rep & 3.380 & 2.794 & 3.402 \\ \cline{2-8}
& \multirow{3}{*}{Auffusion} & \multirow{3}{*}{200} & \multirow{3}{*}{sys8} & Add & 3.890 & 3.396 & 3.454 \\
& & & & Del & 3.871 & 3.552 & 3.420 \\
& & & & Rep & 3.840 & 3.392 & 3.450 \\ \hline
\multirow{18}{*}{\begin{tabular}[c]{@{}c@{}} ZETA\end{tabular}} & \multirow{9}{*}{Tango2} & \multirow{3}{*}{75} & \multirow{3}{*}{sys9} & Add & 3.798 & 3.530 & 3.842 \\
& & & & Del & 3.620 & 3.154 & 3.604 \\
& & & & Rep & 3.556 & 3.306 & 3.478 \\ \cline{3-8}
& & \multirow{3}{*}{125} & \multirow{3}{*}{sys10} & Add & 3.710 & 3.498 & 3.588 \\
& & & & Del & 3.696 & 3.108 & 3.506 \\
& & & & Rep & 3.646 & 3.456 & 3.342 \\ \cline{3-8}
& & \multirow{3}{*}{175} & \multirow{3}{*}{sys11} & Add & 3.638 & 3.518 & 3.480 \\
& & & & Del & 3.758 & 3.658 & 3.424 \\
& & & & Rep & 3.568 & 3.602 & 3.320 \\ \cline{2-8}
& \multirow{9}{*}{Auffusion} & \multirow{3}{*}{75} & \multirow{3}{*}{sys12} & Add & 3.622 & 2.836 & 3.566 \\
& & & & Del & 3.666 & 3.088 & 3.454 \\
& & & & Rep & 3.628 & 3.164 & 3.428 \\ \cline{3-8}
& & \multirow{3}{*}{125} & \multirow{3}{*}{sys13} & Add & 3.682 & 3.180 & 3.482 \\
& & & & Del & 3.688 & 3.522 & 3.460 \\
& & & & Rep & 3.690 & 3.592 & 3.308 \\ \cline{3-8}
& & \multirow{3}{*}{175} & \multirow{3}{*}{sys14} & Add & 3.664 & 3.302 & 3.202 \\
& & & & Del & 3.706 & 3.546 & 3.222 \\
& & & & Rep & 3.636 & 3.448 & 3.078 \\ \hline
\multirow{18}{*}{CycleDiffusion} & \multirow{9}{*}{Tango2} & \multirow{3}{*}{75} & \multirow{3}{*}{sys15} & Add & 3.680 & 3.117 & 3.528 \\
& & & & Del & 3.518 & 2.465 & 3.690 \\
& & & & Rep & 3.580 & 2.960 & 3.567 \\ \cline{3-8}
& & \multirow{3}{*}{125} & \multirow{3}{*}{sys16} & Add & 3.552 & 3.442 & 3.318 \\
& & & & Del & 3.417 & 2.940 & 3.375 \\
& & & & Rep & 3.365 & 3.170 & 3.247 \\ \cline{3-8}
& & \multirow{3}{*}{175} & \multirow{3}{*}{sys17} & Add & 3.240 & 3.268 & 3.020 \\
& & & & Del & 3.283 & 3.038 & 3.075 \\
& & & & Rep & 3.355 & 3.132 & 3.067 \\ \cline{2-8}
& \multirow{9}{*}{Auffusion} & \multirow{3}{*}{75} & \multirow{3}{*}{sys18} & Add & 3.575 & 3.020 & 3.583 \\
& & & & Del & 3.460 & 2.845 & 3.517 \\
& & & & Rep & 3.355 & 2.900 & 3.167 \\ \cline{3-8}
& & \multirow{3}{*}{125} & \multirow{3}{*}{sys19} & Add & 3.305 & 3.035 & 3.065 \\
& & & & Del & 3.385 & 3.145 & 3.125 \\
& & & & Rep & 3.390 & 3.130 & 2.953 \\ \cline{3-8}
& & \multirow{3}{*}{175} & \multirow{3}{*}{sys20} & Add & 3.177 & 2.987 & 2.825 \\
& & & & Del & 3.318 & 3.283 & 2.945 \\
& & & & Rep & 3.290 & 3.207 & 2.907 \\ \hline
\multirow{3}{*}{AP-adapter} & \multirow{3}{*}{Audioldm2} & \multirow{3}{*}{-} & \multirow{3}{*}{sys21} & Add & 3.387 & 2.758 & 2.675 \\
& & & & Del & 3.292 & 2.778 & 2.835 \\
& & & & Rep & 3.380 & 3.035 & 2.450 \\ \hline
\multirow{3}{*}{Musicgen} & \multirow{3}{*}{-} & \multirow{3}{*}{-} & \multirow{3}{*}{sys22} & Add & 3.262 & 2.880 & 2.673 \\
& & & & Del & 3.188 & 3.115 & 2.793 \\
& & & & Rep & 3.175 & 2.910 & 2.615 \\ \hline
\multirow{3}{*}{Human Edit} & \multirow{3}{*}{-} & \multirow{3}{*}{-} & \multirow{3}{*}{sys23} & Add & 3.784 & 3.740 & 3.940 \\
& & & & Del & 3.947 & 4.022 & 4.037 \\
& & & & Rep & 3.937 & 3.891 & 3.991 \\ \hline
\end{tabular}
\label{tab:my-table}
\end{table*}

\newpage

\end{document}